\documentclass[11pt,fleqn]{article}
\usepackage{amsmath,amssymb,graphicx}
\usepackage{xcolor,soul}

\usepackage[colorlinks,citecolor=green!60!black,linkcolor=red!60!black,pagebackref]{hyperref}
\renewcommand*{\backref}[1]{}
\renewcommand*{\backrefalt}[4]{%
\ifcase #1%
\marginpar{\tiny no cite}
\or
 $\rightarrow$~p.~#2.%
\else
  $\rightarrow$~pp.~#2.%
\fi
}

\begin{document}
\sloppy

\newtheorem{axiom}{Axiom}[section]
\newtheorem{conjecture}[axiom]{Conjecture}
\newtheorem{corollary}[axiom]{Corollary}
\newtheorem{definition}[axiom]{Definition}
\newtheorem{example}[axiom]{Example}
\newtheorem{lemma}[axiom]{Lemma}
\newtheorem{observation}[axiom]{Observation}
\newtheorem{proposition}[axiom]{Proposition}
\newtheorem{theorem}[axiom]{Theorem}

\newcommand{\proof}{\emph{Proof.}\ \ }
\newcommand{\qed}{~~$\Box$}

\newcommand{\ppp}{{\cal P}^*}
\newcommand{\ddd}[1]{{\small $d_{#1}$}}

\newcommand{\eee}[1]{E[#1]}
\newcommand{\ees}[1]{E_s[#1]}
\newcommand{\eex}[2]{E_{#1}[#2]}
\newcommand{\eey}[1]{E_{#1}}
\newcommand{\eez}{E_s}

\newcommand{\fff}[1]{F[#1]}
\newcommand{\ffs}[1]{F_s[#1]}
\newcommand{\ffx}[2]{F_{#1}[#2]}
\newcommand{\ffy}[1]{F_{#1}}
\newcommand{\ffz}{F_s}

\title{{\bf The one-dimensional Euclidean domain:\\ Finitely many obstructions are not enough}}
\author{
\sc Jiehua Chen\thanks{{\tt jiehua.chen@tu-berlin.de}.
Department of Software Engineering and Theoretical Computer Science, TU Berlin, Germany}
\and
\sc Kirk R.\ Pruhs\thanks{{\tt kirk@cs.pitt.edu}.
Computer Science Department, University of Pittsburgh, USA}
\and
\sc Gerhard J.\ Woeginger\thanks{{\tt gwoegi@win.tue.nl}. 
Department of Mathematics and Computer Science, TU Eindhoven, Netherlands} 
}
\date{}
\maketitle

\begin{abstract}
We show that one-dimensional Euclidean preference profiles can not be characterized in
terms of finitely many forbidden substructures.
This result is in strong contrast to the case of single-peaked and single-crossing preference 
profiles, for which such finite characterizations have been derived in the literature.

\end{abstract}

\bigskip
\section{Introduction}
Single-peakedness, single-crossingness, and one-dimensional Euclideanness are popular domain 
restrictions that show up in a variety of models in the social sciences and economics.
In many situations, these domain restrictions guarantee the existence of a desirable entity that
would not exist without the restriction, as for instance a  strategy-proof collective choice rule, 
or a Condorcet winner, or an equilibrium point.
\begin{itemize}
\item 
Preferences are \emph{single-peaked}, if there exists a linear ordering of the alternatives such 
that any voter's preference relation along this ordering is either always increasing, always 
decreasing, or first increasing and then decreasing.
\item
Preferences are \emph{single-crossing}, if there exists a linear ordering of the voters such that
for any pair of alternatives along this ordering, there is a single spot where the voters switch
from preferring one alternative above the other one.
\item
Preferences are \emph{one-dimensional Euclidean}, if there exists a common embedding of voters 
and alternatives into the real numbers, such that every voter prefers alternatives that are 
embedded close to him to alternatives that are embedded farther away from him.
\end{itemize}

Single-peakedness goes back to the seminal work of Black~\cite{Black1948}; among many other 
nice consequences, single-peakedness implies transitivity (Inada~\cite{Inada1969}) and 
non-manipulability of the majority rule (Moulin~\cite{Moulin1980}).

Single-crossingness goes back to the work of Karlin \cite{Karlin1968} in applied mathematics;
Mirrlees \cite{Mirrlees1971} and Roberts~\cite{Roberts1977} apply it in the theory of optimal 
income taxation, and Diamond \& Stiglitz \cite{DiSt1974} use it in the economics of uncertainty.
Single-crossingness also plays a role in 
coalition formation (Demange~\cite{Demange1994}; Kung~\cite{Kung2006}),
income redistribution (Meltzer \& Richard~\cite{MeRi1981}),
local public goods and stratification (Westhoff~\cite{Westhoff1977}; Epple \& Platt~\cite{EpPl1998}),
in the choice of constitutional voting rules (Barber\`a \& Jackson~\cite{BaJa2004}),
and in the analysis of the majority rule (Grandmont~\cite{Grandmont1978}; Gans \& Smart~\cite{GaSm1996}). 

One-dimensional Euclidean preference structures go back to Hotelling \cite{Hotelling1929}.
They have been discussed by Coombs~\cite{Coombs1964} under the name `unidimensional unfolding' 
representations, and they unite all the good properties of single-peaked and single-crossing 
preference structures.
Doignon \& Falmagne \cite{DoFa1994} discuss Euclidean preference structures in the context of 
behavioral sciences, and Brams, Jones \& Kilgour \cite{BrJoKi2002} discuss them in the context 
of coalition formation.  

\paragraph{Obstructions.}
The scientific literature contains many characterizations of combinatorial objects in terms 
of forbidden substructures or \emph{obstructions}.
For instance, Kuratowski's theorem~\cite{Kuratowski1930} characterizes planar graphs in terms of two 
obstructions: a graph is planar if and only if it does not contain a subdivided $K_5$ or $K_{3,3}$.
In a similar spirit, Lekkerkerker \& Boland \cite{LeBo1962} characterize interval graphs through five
(infinite) families of forbidden induced subgraphs, and F\"oldes \& Hammer characterize split graphs 
in terms of three forbidden induced subgraphs.
Hoffman, Kolen \& Sakarovitch~\cite{HoKoSa1985} characterize totally-balanced $0$-$1$-matrices
in terms of certain forbidden submatrices.
The characterizations of split graphs and totally-balanced $0$-$1$-matrices use a finite 
number of obstructions, while the characterizations of planar graphs and interval graphs both 
involve infinitely many obstructions.
In the area of social choice, Ballester \& Haeringer~\cite{BaHa2011} characterize single-peaked 
preference profiles and group-separable preference profiles in terms of a small finite number of 
obstructions.
Also single-crossing preference profiles allow a characterization by finitely many obstructions;
see Bredereck, Chen \& Woeginger~\cite{BrChWo2013a}.

A characterization by finitely many obstructions has many positive consequences.
Whenever a family $\cal F$ of combinatorial objects allows such a finite characterization, this 
directly implies the existence of a polynomial time algorithm for recognizing the members of 
$\cal F$: one may simply work through the obstructions one by one, and check whether the 
considered object contains the obstruction. 
By looking deeper into the combinatorial structure of such families $\cal F$, one usually 
manages to find recognition algorithms that are much faster than this simple approach.
As an example, there exist sophisticated algorithms for recognizing single-peaked preference 
profiles that are due to 
Bartholdi \& Trick~\cite{BaTr1986},
Doignon \& Falmagne~\cite{DoFa1994}, and 
Escoffier, Lang \& {\"O}zt{\"u}rk~\cite{EsLaOz2008}.
Also single-crossingness can be recognized very efficiently; see
Doignon \& Falmagne~\cite{DoFa1994},
Elkind, Faliszewski \& Slinko~\cite{ElFaSl2012}, and
Bredereck, Chen \& Woeginger~\cite{BrChWo2013a}.

As another positive consequence, a characterization by finitely many obstructions often helps 
us in understanding the algorithmic and combinatorial behavior of family $\cal F$.
For example, Bredereck, Chen \& Woeginger~\cite{BrChWo2013b} investigate the problem of
deciding whether a given preference profile is close to a nicely structured preference profile.
The distance is measured by the number of voters or alternatives that have to be deleted from
the given profile so as to reach a nicely structured profile.
For the cases where `nicely structured' means single-peaked or single-crossing, the proofs 
in~\cite{BrChWo2013b} are heavily based on characterizations \cite{BaHa2011,BrChWo2013a} by 
finitely many obstructions.
Elkind \& Lackner \cite{ElLa2014} study similar questions and derive approximation algorithms 
for the number of deleted voters or alternatives.
All results in \cite{ElLa2014} are centered around preference profiles that can be 
characterized by a finite number of obstructions, and some of the theorems are parameterized 
by the obstruction set.

\paragraph{Scope and contribution of this paper.}
As the one-dimensional Euclidean profiles form a special case of single-peaked and 
single-crossing profiles (see Section~\ref{ssec:Euclidean} for more information on this), 
every obstruction to single-peakedness and every obstruction to single-crossingness 
will automatically also form an obstruction to one-dimensional Euclideanness.
Now the question arises: \emph{``Are there any further obstructions to one-dimensional 
Euclideanness?''}
To which Clyde Coombs~\cite{Coombs1964} answered back in 1964: 
\emph{``Yes, there are!''} (again, see Section~\ref{ssec:Euclidean} for more information).
This immediately takes us to another question: \emph{``Is there a characterization 
of one-dimensional Euclideanness in terms of finitely many obstructions?''}
The answer to this second question is negative, as we are going to show in this paper.

To this end, we construct an infinite sequence of preference profiles that satisfy two 
crucial properties.
First, none of these profiles is one-dimensional Euclidean. 
Secondly, every profile just \emph{barely} violates one-dimensional Euclideanness, as
the deletion of an arbitrary voter immediately makes the profile one-dimensional Euclidean.
The second property implies that each profile in the sequence is on the edge of being 
Euclidean, and that the reason for its non-Euclideanness must lie in its overall structure.
In other words each of these infinitely many profiles yields a separate obstruction for 
one-dimensional Euclideanness, and this is exactly what we want to establish.

The definition of the infinite profile sequence and the resulting analysis are quite involved.
Ironically, the complexity of our proof is a consequence of the very statement we are going 
to prove.
As part of our proof, we have to argue that the deletion of an arbitrary voter from an
arbitrary profile in the sequence yields a one-dimensional Euclidean profile.
Now if there was a characterization of one-dimensional Euclideanness by finitely many 
obstructions, then this argument would be relatively easy to get through: we could 
simply analyze the preference structure and show that the deletion of any voter 
removes all obstructions.
But unfortunately, such a characterization does not exist.
The only viable (and fairly tedious) approach is to explicitly specify the corresponding 
Euclidean representations (one representation per deleted voter!) and to prove by case 
distinctions that each such representation correctly encodes the preferences of all the 
remaining voters.

\paragraph{Organization of the paper.}
In Section~\ref{sec:defi} we summarize the central definitions, state useful observations, 
and provide some examples.
In Section~\ref{sec:main} we formulate our main results in Theorems~\ref{th:main.1} and
\ref{th:main.2}, and we show how Theorem~\ref{th:main.2} follows from Theorem~\ref{th:main.1}.
The five Sections~\ref{sec:profile} through~\ref{sec:correct} present the long and technical 
proof of Theorem~\ref{th:main.1}.
Section~\ref{sec:conclusion} completes the paper with a short discussion.  

\section{Definitions, notations, and examples}
\label{sec:defi}
Let $1,\ldots,m$ be $m$ alternatives and let $v_1,\ldots,v_n$ be $n$ voters.
A \emph{preference profile} specifies the \emph{preference orderings} of the voters, where voter $v_i$ 
ranks the alternatives according to a strict linear order $\succ_i$.
For alternatives $a$ and $b$, the relation $a\succ_i b$ means that voter $v_i$ strictly prefers $a$ to $b$.
If the meaning is clear from the context, we will sometimes simply write $\succ$ instead of $\succ_i$ 
and suppress the dependence on $i$. 
A profile with $n$ voters and $m$ alternatives will be called an $n\times m$ profile.

\subsection{Single-peaked profiles}
\label{ssec:single-peaked}
A linear ordering of the alternatives is single-peaked with respect to a fixed voter $v_i$,
if the preferences of $v_i$ taken along this ordering have a single local maximum.
A preference profile is \emph{single-peaked}, if it allows an ordering of the 
alternatives that is single-peaked with respect to every voter.  

Note that for every single-peaked permutation $\pi(1),\pi(2),\ldots,\pi(m)$ of the 
alternatives, also the reverse permutation $\pi(m),\ldots,\pi(2),\pi(1)$ is single-peaked.
The following proposition states a characterization of single-peakedness in terms
of finitely many obstructions.

\begin{proposition}
\label{pr:peaked}
(Ballester \& Haeringer~\cite{BaHa2011})\\
A preference profile is single-peaked, if and only if it avoids the following two obstructions.
The first obstruction is a $3\times3$ profile with alternatives $a,b,c$:
\begin{quote}
Voter~ $v_1$:~ $\{b,c\}\succ_1 a$ \\[0.5ex]
Voter~ $v_2$:~ $\{a,c\}\succ_2 b$ \\[0.5ex]
Voter~ $v_3$:~ $\{a,b\}\succ_3 c$ 
\end{quote}
The second obstruction is a $2\times4$ profile with alternatives $a,b,c,d$:
\begin{quote}
Voter~ $v_1$:~ $a\succ_1 b\succ_1 c$ ~and~ $d\succ_1 b$ \\[0.5ex]
Voter~ $v_2$:~ $c\succ_2 b\succ_2 a$ ~and~ $d\succ_2 b$ 
\end{quote}
\end{proposition}

\subsection{Single-crossing profiles}
\label{ssec:single-crossing}
A linear ordering of the voters is single-crossing with respect to two alternatives $a$ and $b$, 
if the ordered list of voters can be partitioned into an initial piece and a final piece 
such that all voters in the initial piece have the same relative ranking of $a$ and $b$,
while all voters in the final piece rank them in the opposite way.
A preference profile is \emph{single-crossing}, if it allows an ordering of the voters that 
is single-crossing with respect to every possible pair of alternatives.  

The following proposition states a characterization of single-crossingness in terms
of finitely many obstructions.

\begin{proposition}
\label{pr:crossing}
(Bredereck, Chen \& Woeginger~\cite{BrChWo2013a})\\
A preference profile is single-crossing, if and only if it avoids the following two obstructions.
The first obstruction is a $3\times6$ profile with (not necessarily distinct) alternatives $a,b,c,d,e,f$:
\begin{quote}
Voter~ $v_1$:~ $b\succ_1 a$ ~and~ $c\succ_1 d$ ~and~ $e\succ_1 f$\\[0.5ex]
Voter~ $v_2$:~ $a\succ_2 b$ ~and~ $d\succ_2 c$ ~and~ $e\succ_2 f$\\[0.5ex]
Voter~ $v_3$:~ $a\succ_3 b$ ~and~ $c\succ_3 d$ ~and~ $f\succ_3 e$
\end{quote}
The second obstruction is a $4\times4$ profile with (not necessarily distinct) alternatives $a,b,c,d$:
\begin{quote}
Voter~ $v_1$:~ $a\succ_1 b$ ~and~ $c\succ_1 d$\\[0.5ex]
Voter~ $v_2$:~ $a\succ_2 b$ ~and~ $d\succ_2 c$\\[0.5ex]
Voter~ $v_3$:~ $b\succ_3 a$ ~and~ $c\succ_3 d$\\[0.5ex]
Voter~ $v_4$:~ $b\succ_4 a$ ~and~ $d\succ_4 c$
\end{quote}
\end{proposition}

\subsection{One-dimensional Euclidean profiles}
\label{ssec:Euclidean}
Consider a common embedding of the voters and alternatives into the real number line, that 
assigns to every alternative $j$ a real number $\eee{j}$ and that assigns to every voter $v_i$ 
a real number $\fff{i}$.
A preference profile is \emph{one-dimensional Euclidean}, if there exists such a common 
\emph{Euclidean representation} of the voters and alternatives, such that for every voter $v_i$ 
and for every pair $a$ and $b$ of alternatives, $a\succ_i b$ holds if and only if the distance 
from $\fff{i}$ to $\eee{a}$ is strictly smaller than the distance from $\fff{i}$ to $\eee{b}$.
In other words, small spatial distances from the point $\fff{i}$ indicate strong preferences 
of voter $v_i$.

It is well-known (and easy to see) that every one-dimensional Euclidean profile is simultaneously 
single-peaked and single-crossing:
the left-to-right ordering of the alternatives along the Euclidean representation is single-peaked, and
the left-to-right ordering of the voters along the Euclidean representation is single-crossing.
Coombs~\cite[page 91]{Coombs1964} discusses a $16\times6$ preference profile that is both
single-peaked and single-crossing, but fails to be one-dimensional Euclidean.
The following example contains the smallest profile known to us that has these intriguing properties. 

\begin{example}
\label{ex:3x6}
Consider the following $3\times6$ profile $\cal P$ (for the sake of readability, the preference 
orders are simply listed left to right from most preferred to least preferred alternative): 
\begin{quote}
Voter~ $v_1$:~~ 3~ 2~ 1~ 4~ 5~ 6 \\[0.5ex]
Voter~ $v_2$:~~ 3~ 4~ 2~ 5~ 6~ 1 \\[0.5ex]
Voter~ $v_3$:~~ 5~ 4~ 3~ 6~ 2~ 1 
\end{quote}
This profile $\cal P$ is single-peaked with respect to the ordering $1,2,3,4,5,6$ of alternatives,
and it is single-crossing with respect to the ordering $v_1,v_2,v_3$ of voters.
Furthermore it can be shown by case distinctions that $\cal P$ is not one-dimensional Euclidean.
\end{example}

As the profile in Example~\ref{ex:3x6} is single-peaked and single-crossing, it does not contain
any of the obstructions listed in Propositions~\ref{pr:peaked} and~\ref{pr:crossing}.
Hence there must be some other obstruction contained in it, that is responsible for its
non-Euclideanness.
Example~\ref{ex:3x6} and the $16\times6$ profile of Coombs provide first indications that the 
obstructions for one-dimensional Euclideanness might be complex and intricate to analyze.

The following two propositions state simple observations that will be used repeatedly in our arguments.

\begin{proposition}
\label{pr:Euclidean.1}
Let $a$ and $b$ be two alternatives in a Euclidean embedding $(E,F)$ of some profile 
with $\eee{a}<\eee{b}$.
Then voter $v_i$ prefers $a$ to~$b$ if and only if $\fff{i}<\frac12(\eee{a}+\eee{b})$,
and he prefers $b$ to~$a$ if and only if $\fff{i}>\frac12(\eee{a}+\eee{b})$.
\end{proposition}

\begin{proposition}
\label{pr:Euclidean.2}
Let $a,b,c$ be three alternatives in a Euclidean embedding $(E,F)$ of some preference 
profile with $\eee{a}<\eee{b}<\eee{c}$.

$\bullet$~ If voter $v_i$ prefers $a\succ_i b$, then he also prefers $b\succ_i c$.

$\bullet$~ If voter $v_i$ prefers $c\succ_i b$, then he also prefers $b\succ_i a$.
\end{proposition}

Finally, we mention that the (mathematical) literature on one-dimensional Euclidean preference 
profiles is scarce.
Doignon \& Falmagne \cite{DoFa1994} and Knoblauch \cite{Knoblauch2010} design polynomial time
algorithms for deciding whether a given preference profile has a one-dimensional Euclidean
representation.
The approaches in \cite{DoFa1994,Knoblauch2010} are not purely combinatorial, as they are
partially based on linear programming formulations.

\section{Statement of the main results}
\label{sec:main}
In this section we formulate the two (closely related) main results of this paper.
The first result is technical and states the existence of infinitely many non-Euclidean 
profiles that are minimal with respect to voter deletion.

\begin{theorem}
\label{th:main.1}
For any integer $k\ge2$, there exists a preference profile $\ppp_k$ with $n=2k$ voters
and $m=4k$ alternatives, such that the following holds.
\begin{itemize}
\itemsep=0.0ex
\item[(a)] Profile $\ppp_k$ is not one-dimensional Euclidean. 
\item[(b)] Profile $\ppp_k$ is minimal in the following sense: the deletion of an arbitrary 
voter from $\ppp_k$ yields a one-dimensional Euclidean profile.
\end{itemize}
\end{theorem}

The proof of Theorem~\ref{th:main.1} is long and will fill most of the rest of this paper.
Here is a quick overview of this proof:
Section~\ref{sec:profile} describes the profiles $\ppp_k$.
Section~\ref{sec:Euclid} shows that every profile $\ppp_k$ satisfies property~(a) in Theorem~\ref{th:main.1},
while the three Sections~\ref{sec:embedding} through~\ref{sec:correct} establish property~(b).
Section~\ref{sec:embedding} defines the underlying Euclidean representations, 
Section~\ref{sec:technical} lists a number of technical auxiliary statements, and
Section~\ref{sec:correct} establishes the correctness of the Euclidean representations.

As an immediate consequence of Theorem~\ref{th:main.1}, we derive our second main result 
(which essentially repeats the title of the paper) in the following theorem.

\begin{theorem}
\label{th:main.2}
One-dimensional Euclidean preference profiles can not be characterized in terms of finitely many obstructions.
\end{theorem}
\proof
Suppose for the sake of contradiction that such a characterization with finitely many obstructions
would exist.
Let $t$ denote the largest number of voters in any obstruction, and consider a profile $\ppp_k$ 
from Theorem~\ref{th:main.1} with $k\ge t$.
As $\ppp_k$ is not one-dimensional Euclidean by property (a), it must contain one of these finitely 
many obstructions with at most $t$ voters.
Fix such an obstruction.
As profile $\ppp_k$ contains $2k>t+1$ voters, one of its voters is not involved in the obstruction.
If we delete this voter, the resulting profile will still contain the obstruction; hence it is not 
one-dimensional Euclidean, which contradicts property~(b).
\qed

\section{Definition of the profiles}
\label{sec:profile}
In this section we start the proof of Theorem~\ref{th:main.1} by defining the underlying profiles $\ppp_k$.
The properties (a) and (b) stated in Theorem~\ref{th:main.1} will be established in the following sections.

We consider $n=2k$ voters called $v_1,v_2,\ldots,v_{2k}$ together with $m=4k$ alternatives called 
$1,2,3,\ldots,4k$. 
The preference orderings of the voters will be pasted together from the following preference 
pieces $X_i$, $Y_i$, $Z_i$ with $1\le i\le k$.
\begin{align*}
X_i &~:=~~ {2k+2i-2} ~\succ~ {2k+2i-3} ~\succ~ {2k+2i-4} ~\succ~ \ldots ~\succ~ {2i+2} \\[0.5ex]
Y_i &~:=~~ {2i-2}    ~\succ~ {2i-3}    ~\succ~ {2i-4}    ~\succ~ \ldots ~\succ~ {1}    \\[0.5ex]
Z_i &~:=~~ {2k+2i+1} ~\succ~ {2k+2i+2} ~\succ~ {2k+2i+3} ~\succ~ \ldots ~\succ~ {4k} 
\end{align*}
Note that for every $i=1,\ldots,k$, the corresponding three pieces $X_i$, $Y_i$, $Z_i$ cover contiguous 
intervals of respectively $2k-3$, $2i-2$, $2k-2i$ alternatives.
Hence these three pieces jointly cover $4k-5$ of the alternatives, and only the five alternatives in
the set 
\[ U_i ~=~ \{2i-1,~ 2i,~ 2i+1\} ~\cup~ \{2k+2i-1,~ 2k+2i\}\] 
remain uncovered. 
Note furthermore that the pieces $Y_1$ and $Z_k$ are empty.
Now let us define the preference orderings of the voters.
The two voters $v_{2i-1}$ and $v_{2i}$ always form a couple with fairly similar preferences.
For $1\le i\le k-1$, these voters $v_{2i-1}$ and $v_{2i}$ have the following preferences:
\begin{subequations}
\begin{alignat}{3}
\label{eq:hua.odd} v_{2i-1}:&\quad X_i\succ{2i+1}   \succ{2k+2i-1}\succ{2i}  \succ{2i-1}\succ{2k+2i}\succ Y_i\succ Z_i\\[0.5ex]
\label{eq:hua.even}v_{2i}:  &\quad X_i\succ{2k+2i-1}\succ{2k+2i}  \succ{2i+1}\succ{2i}  \succ{2i-1} \succ Y_i\succ Z_i.  
\end{alignat}
\end{subequations}
Note that the voters $v_{2i-1}$ and $v_{2i}$ both rank the three alternatives $2i+1$, $2i$, $2i-1$ 
in $U_i$ in the same decreasing order, with the two other alternatives $2k+2i-1$ and $2k+2i$ 
shuffled into that order.
The last two voters $v_{2k-1}$ and $v_{2k}$ are defined separately:
\begin{subequations}
\begin{alignat}{3}
\label{eq:hua.2k-1} v_{2k-1}:&\quad X_k\succ {2k+1}\succ {4k-1}\succ {2k}\succ {2k-1}  \succ {4k} \succ Y_k \\[0.5ex]
\label{eq:hua.2k}   v_{2k}:  &\quad X_k\succ {2k+1}\succ {2k}\succ \ldots\ldots \succ{3} \succ{2} \succ{4k-1} \succ{4k} \succ{1}
\end{alignat}
\end{subequations}
Since piece $Z_k$ is empty, the preferences of voter $v_{2k-1}$ in \eqref{eq:hua.2k-1} 
actually run in parallel with the preferences of the other odd-index voters $v_{2i-1}$ 
with $1\le i\le k-1$ in \eqref{eq:hua.odd}.
The last voter~$v_{2k}$, however, behaves very differently from the other even-index voters: 
on top of his preference list are the alternatives in piece $X_k$, followed by an intermingling 
of the alternatives in piece $Y_k$ and set $U_k$ (first the alternatives $2k+1,\ldots,2$ in 
decreasing order, and then the three alternatives $4k-1$, $4k$, and $1$).

\begin{example}
\label{ex:part1}
For $k=4$, the preference profile $\ppp_4$ has $n=8$ voters and $m=16$ alternatives and looks as follows
(all preference orders are listed left to right from most preferred to least preferred alternative):
\begin{center}
\begin{tabular}{|l|ccccc|ccccccccccc|}
\hline
 $v_1:$ & 8 & 7 & 6 & 5 & 4 & 3 & 9 &2 &1 &10 &\multicolumn{1}{|c}{11} &12 &13 &14 &15 &16 \\[0.6ex]
 $v_2:$ & 8 & 7 & 6 & 5 & 4 & 9 &10 &3 &2 & 1 &\multicolumn{1}{|c}{11} &12 &13 &14 &15 &16 \\[0.6ex] \cline{12-13}
 $v_3:$ &10 & 9 & 8 & 7 & 6 & 5 &11 &4 &3 &12 &\multicolumn{1}{|c}{ 2} & 1 &\multicolumn{1}{|c}{13} &14 &15 &16 \\[0.6ex]
 $v_4:$ &10 & 9 & 8 & 7 & 6 &11 &12 &5 &4 & 3 &\multicolumn{1}{|c}{ 2} & 1 &\multicolumn{1}{|c}{13} &14 &15 &16 \\[0.6ex] \cline{14-15}
 $v_5:$ &12 &11 &10 & 9 & 8 & 7 &13 &6 &5 &14 &\multicolumn{1}{|c}{ 4} & 3 & 2 & 1 &\multicolumn{1}{|c}{15} &16 \\[0.6ex]
 $v_6:$ &12 &11 &10 & 9 & 8 &13 &14 &7 &6 & 5 &\multicolumn{1}{|c}{ 4} & 3 & 2 & 1 &\multicolumn{1}{|c}{15} &16 \\[0.6ex] \cline{16-17}
 $v_7:$ &14 &13 &12 &11 &10 & 9 &15 &8 &7 &16 &\multicolumn{1}{|c}{ 6} & 5 & 4 & 3 & 2 & 1 \\[0.3ex] \cline{7-17}
 $v_8:$ &14 &13 &12 &11 &10 & 9 & 8 &7 &6 & 5 & 4 & 3 & 2 &15 &16 & 1 \\[0.6ex]
\hline
\end{tabular} 
\end{center}
The alternatives in the five leftmost columns form the pieces $X_i$. 
In the first seven rows, the five middle columns correspond to the sets $U_i$, while the 
remaining six columns belong to pieces $Y_i$ and $Z_i$.
The last row illustrates the extraordinary behavior of the last voter $v_8$.
\qed
\end{example}

\section{The profiles are not Euclidean}
\label{sec:Euclid}
In this section, we will discuss single-crossing, single-peaked and one-dimensional
Euclidean properties of the profiles $\ppp_k$.
First, it can readily be seen that every profile $\ppp_k$ with $k\ge2$ is single-crossing 
with respect to the ordering $v_1,v_2,\ldots,v_{2k-2},v_{2k},v_{2k-1}$ of the voters
(that is, the natural ordering of voters by increasing index, but with the last two voters 
$v_{2k-1}$ and $v_{2k}$ swapped). 
As this single-crossing property is of no relevance for our further considerations, 
the simple proof is omitted.
Next, let us turn to single-peakedness.

\begin{lemma}
\label{le:peaked}
For $k\ge2$, the profile $\ppp_k$ is single-peaked.
Furthermore, the only two single-peaked orderings of the alternatives are the 
increasing ordering $1,2,3,\ldots,4k$ and the decreasing ordering $4k,\ldots,3,2,1$.
\end{lemma}
\proof
Every voter $v_{2i-1}$ and $v_{2i}$ with $1\le i\le k$ has alternative $2k+2i-2$ as his top preference.
Furthermore, he ranks the small alternatives $1,2,\ldots,2k+2i-2$ decreasingly and he ranks the large
alternatives $2k+2i-2,2k+2i-1,\ldots,4k$ increasingly. 
Hence $\ppp_k$ indeed is single-peaked with respect to $1,2,3,\ldots,4k$ and $4k,\ldots,3,2,1$.

Next consider an arbitrary single-peaked permutation $\pi(1),\pi(2),\ldots,\pi(4k)$ of the alternatives.
Since $4k$ and $1$ are the least preferred choices of voters $v_1$ and $v_{2k}$, these two alternatives 
must be extremal in the single-peaked ordering; by symmetry we will assume $\pi(1)=1$ and $\pi(4k)=4k$.
\begin{itemize}
\item Voter $v_1$ ranks $1\succ2k+2\succ2k+3\succ\ldots\succ4k$, without other alternatives ranked
inbetween.
This implies $\pi(x)=x$ for $2k+2\le x\le4k$.
\item Voter $v_{2k-1}$ ranks $2k+2\succ2k+1\succ2k\succ\ldots\succ3\succ2\succ1$.
This now implies $\pi(x)=x$ also for the alternatives $x$ with $1\le x\le 2k+1$.
\end{itemize}
Summarizing, we have $\pi(x)=x$ for all $x$, and this completes the proof.
\qed

\bigskip
The following lemma shows that every profile $\ppp_k$ satisfies property (a) of Theorem~\ref{th:main.1}.

\begin{lemma}
\label{le:main.(a)}
For $k\ge2$, the profile $\ppp_k$ is not one-dimensional Euclidean.
\end{lemma}
\proof
We suppose for the sake of contradiction that profile $\ppp_k$ is one-dimensional Euclidean.
Let $\fff{j}$ for $j=1,\ldots,2k$ and $\eee{i}$ for $i=1,\ldots,4k$ denote a corresponding Euclidean
representation of the voters and alternatives.
As the Euclidean representation induces a single-peaked ordering of the alternatives, we will 
assume by Lemma~\ref{le:peaked} that the alternatives are embedded in increasing order with
\begin{equation}
\label{eq:chain}
\eee{1} ~<~ \eee{2} ~<~ \eee{3} ~<~ \ldots ~<~ \eee{4k-1} ~<~ \eee{4k}.
\end{equation}
Next, we claim that in any Euclidean representation under \eqref{eq:chain}, the embedded 
alternatives satisfy the following system of inequalities:
\begin{subequations}
\begin{alignat}{3}
\eee{2k+2i-1}+\eee{2i}  &~<~ &&\eee{2k+2i}+\eee{2i-1}   &&\mbox{\qquad for $1\le i\le k$}   \label{eq:cyc.1} \\[0.5ex]
\eee{2k+2i}+\eee{2i+1}  &~<~ &&\eee{2k+2i-1}+\eee{2i+2} &&\mbox{\qquad for $1\le i\le k-1$} \label{eq:cyc.2} \\[0.5ex]
\eee{4k}+\eee{1}        &~<~ &&\eee{4k-1}+\eee{2}     \label{eq:cyc.3}
\end{alignat}
\end{subequations}
The correctness of this system can be seen as follows.
For each $i=1,\ldots,k$, voter $v_{2i-1}$ ranks ${2k+2i-1}\succ {2i}$ and ${2i-1}\succ {2k+2i}$, 
which by Proposition~\ref{pr:Euclidean.1} yields
\[ \frac12\left(\eee{2k+2i-1}+\eee{2i}\right) ~<~ \fff{2i-1} ~<~ \frac12\left(\eee{2k+2i}+\eee{2i-1}\right), \]
which in turn implies \eqref{eq:cyc.1}.
Similarly, for $i=1,\ldots,k-1$ voter $v_{2i}$ ranks ${2i+2}\succ{2k+2i-1}$ and ${2k+2i}\succ{2i+1}$ 
which leads to \eqref{eq:cyc.2}.
Finally, voter $v_{2k}$ ranks ${2}\succ{4k-1}$ and ${4k}\succ{1}$, which implies \eqref{eq:cyc.3}.
This establishes correctness of the system \eqref{eq:cyc.1}--\eqref{eq:cyc.3}.
By adding up all the inequalities in \eqref{eq:cyc.1}--\eqref{eq:cyc.3}, we derive the contradiction
$\sum_{x=1}^{4k}\eee{x}<\sum_{x=1}^{4k}\eee{x}$.
\qed

\section{Definition of the Euclidean representations}
\label{sec:embedding}
In this section, we fix an integer $s$ with $1\le s\le2k$ and construct corresponding
Euclidean embeddings $\ffz$ and $\eez$ of the voters and alternatives in profile $\ppp_k$.
We start by defining the Euclidean embedding $\eez$ of the alternatives.
We anchor the embedding by placing the first alternative at the position
\begin{equation}
\label{eq:alter.0}
\ees{1} ~=~ 0.
\end{equation}
The remaining values $\ees{2},\ldots,\ees{4k}$ are described recursively in 
equations \eqref{eq:alter.1}--\eqref{eq:alter.6} below.
For $1\le i\le k-1$ we set
\begin{equation}
\label{eq:alter.1}
\ees{2i+1}-\ees{2i} ~=~2
\end{equation}
and for $1\le i\le k$ we set
\begin{equation}
\label{eq:alter.2}
\ees{2i}-\ees{2i-1} ~=~ (4i-2s-3 \bmod 4k).
\end{equation}
Note that the relations \eqref{eq:alter.0}--\eqref{eq:alter.2} determine $\ees{x}$ for all $x\le2k$.
For $1\le i\le k-1$ we set
\begin{eqnarray}
\lefteqn{\ees{2k+2i-1}-\ees{2k+2i-2}} \nonumber \\[0.5ex] 
&=& \left\{
    \begin{array}[h]{ll}
      \ees{2k+2i-3}-\ees{2i+1}+2~~&\mbox{if $s\ne2i-1$}\\[0.5ex]
      \ees{2k+2i-3}-\ees{2i+2}+2  &\mbox{if $s=2i-1$.}
    \end{array}
  \right.
\label{eq:alter.3}
\end{eqnarray}
For $1\le i\le k-1$ we define
\begin{equation}
\label{eq:alter.4}
\ees{2k+2i}-\ees{2k+2i-1} ~=~ (4i-2s-1 \bmod 4k).
\end{equation}
Note that the relations \eqref{eq:alter.3} and \eqref{eq:alter.4} determine $\ees{x}$ for 
all $x$ with $2k+1\le x\le 4k-2$.
Finally, we determine the Euclidean embedding of the last two alternatives by defining
\begin{equation}
\label{eq:alter.5}
\ees{4k-1}-\ees{4k-2} ~=~ \left\{
    \begin{array}[h]{ll}
      \ees{4k-3}-\ees{2}+2    &\mbox{if $s\ne2k$}\\[0.5ex]
      \ees{4k-3}-\ees{2k+1}+2 &\mbox{if $s=2k$}
    \end{array}
  \right.
\end{equation}
and
\begin{equation}
\label{eq:alter.6}
\ees{4k}-\ees{4k-1} ~=~ \left\{
    \begin{array}[h]{ll}
       \ees{2}  -\ees{1}-2 &\mbox{if $s\ne2k$}\\[0.5ex]
       \ees{2k+1}-\ees{2k-1} &\mbox{if $s=2k$.}
    \end{array}
  \right.
\end{equation}
This completes the description of the Euclidean embedding $\eez$ of the alternatives.
Note that $\ees{x}$ is integer for all alternatives $x$.

\begin{lemma}
\label{le:order}
The embedding $\eez$ satisfies $\ees{x}<\ees{y}$ for all alternatives $x$ and $y$ with $1\le x<y\le4k$.
In other words, $\eez$ satisfies the chain of inequalities in \eqref{eq:chain}.
\end{lemma}
\proof
The statement follows from \eqref{eq:alter.0}--\eqref{eq:alter.6} by an easy inductive argument.
The right hand sides in \eqref{eq:alter.1}, \eqref{eq:alter.2} and \eqref{eq:alter.4} are all positive. 
The right hand sides in \eqref{eq:alter.3} and \eqref{eq:alter.5} can be seen to be positive by induction.
Finally for $i=1$ and $s\ne2k$, the right hand side of \eqref{eq:alter.2} is a positive odd integer
strictly greater than~$1$; this yields $\ees{2}\ge3$ so that also the right hand side $\ees{2}-\ees{1}-2$ in
\eqref{eq:alter.6} is positive.
\qed

\begin{example}
\label{ex:part2}
We continue our discussion of the profile $\ppp_4$ from Example~\ref{ex:part1}.
For every embedding $\eez$ with $1\le s\le8$, the corresponding row in Table~\ref{tab:distances} 
lists the distances $d_i=\ees{i}-\ees{i-1}$ between pairs of consecutive alternatives according 
to formulas \eqref{eq:alter.1}--\eqref{eq:alter.6}.
For instance the crossing of the row $\eey{5}$ and the column labeled $d_4$ contains an entry 
with value $11$; this means that in the Euclidean representation $\eey{5}$, the distance 
$\eex{5}{4}-\eex{5}{3}$ between the embedded alternatives $3$ and $4$ equals $11$. 
As $\ees{1}=0$, we see that for $2\le i\le4k$ the value $\ees{i}$ then equals $d_2+d_3+\cdots+d_i$.
For instance in $\eey{5}$, alternative $4$ will be embedded in the point $\eex{5}{4}=7+2+11=20$.

The reader will notice that part of the data in Table~\ref{tab:distances} carries a periodic structure. 
For instance every even-indexed column (except the last one) contains a circular shift of the eight 
numbers 1, 3, 5, 7, 9, 11, 13, 15 presented in boldface, which results from formulas \eqref{eq:alter.2} 
and \eqref{eq:alter.4}.
Furthermore, all the entries in the three columns $d_3$, $d_5$, $d_7$ have the same value~$2$ 
according to \eqref{eq:alter.1}.
The numbers in other parts of the table look somewhat irregular and chaotic, which is caused 
by formula \eqref{eq:alter.3}.
For us, the most convenient way of working with this data is via the recursive 
definitions \eqref{eq:alter.0}--\eqref{eq:alter.6}. 
\qed
\end{example}

\begin{table}[tbh]
\begin{center}
\begin{tabular}{|l|ccccccccccccccc|}
\hline
&\ddd{2} &\ddd{3}  &\ddd{4} &\ddd{5}  &\ddd{6} &\ddd{7} &\ddd{8} 
&\ddd{9} &\ddd{10} &\ddd{11}&\ddd{12} &\ddd{13}&\ddd{14}&\ddd{15} &\ddd{16} \\[0.9ex]
\hline
 $\eey{1}$ &{\bf15} &2 &{\bf 3} &2 &{\bf 7} &2 &{\bf11} &13 &{\bf 1} &35 &{\bf 5} &62 &{\bf 9} &145 &{\bf13}\\[0.9ex]
 $\eey{2}$ &{\bf13} &2 &{\bf 1} &2 &{\bf 5} &2 &{\bf 9} &12 &{\bf15} &30 &{\bf 3} &68 &{\bf 7} &151 &{\bf11}\\[0.9ex]
 $\eey{3}$ &{\bf11} &2 &{\bf15} &2 &{\bf 3} &2 &{\bf 7} &24 &{\bf13} &35 &{\bf 1} &81 &{\bf 5} &187 &{\bf 9}\\[0.9ex]
 $\eey{4}$ &{\bf 9} &2 &{\bf13} &2 &{\bf 1} &2 &{\bf 5} &20 &{\bf11} &30 &{\bf15} &68 &{\bf 3} &171 &{\bf 7}\\[0.9ex]
 $\eey{5}$ &{\bf 7} &2 &{\bf11} &2 &{\bf15} &2 &{\bf 3} &32 &{\bf 9} &54 &{\bf13} &97 &{\bf 1} &242 &{\bf 5}\\[0.9ex]
 $\eey{6}$ &{\bf 5} &2 &{\bf 9} &2 &{\bf13} &2 &{\bf 1} &28 &{\bf 7} &46 &{\bf11} &84 &{\bf15} &207 &{\bf 3}\\[0.9ex]
 $\eey{7}$ &{\bf 3} &2 &{\bf 7} &2 &{\bf11} &2 &{\bf15} &24 &{\bf 5} &54 &{\bf 9} &100&{\bf13} &233 &{\bf 1}\\[0.9ex]
 $\eey{8}$ &{\bf 1} &2 &{\bf 5} &2 &{\bf 9} &2 &{\bf13} &20 &{\bf 3} &46 &{\bf 7} &84 &{\bf11} &142 &{\bf33}\\[0.9ex] 
\hline
\end{tabular}
\end{center}
\caption{This table is discussed in Example~\protect{\ref{ex:part2}} and illustrates 
the Euclidean embedding of the alternatives in profile $\ppp_4$. 
Every row is labeled by a corresponding embedding $\eez$.
If a column is labeled by $d_i$, then its entries indicate the Euclidean distances 
$\ees{i}-\ees{i-1}$ between the two consecutively embedded alternatives $i-1$ and $i$.}
\label{tab:distances}
\end{table}

Now let us turn to the Euclidean embedding of the voters.
The Euclidean position $\ffs{j}$ of every voter $v_j$ will be the average of exactly four 
embedded alternatives.
For $1\le i\le k-1$ we define
\begin{equation}
\label{eq:voter.1}
\ffs{2i-1} ~=~ \frac14\left( \ees{2i-1}+\ees{2i}+\ees{2k+2i-1}+\ees{2k+2i}\right).
\end{equation}
Similarly, for $1\le i\le k-1$ we define
\begin{equation}
\label{eq:voter.2}
\ffs{2i} ~=~ \frac14\left( \ees{2i+1}+\ees{2i+2}+\ees{2k+2i-1}+\ees{2k+2i}\right).
\end{equation}
If $s\ne2k$ then voter $v_{2k-1}$ is embedded according to \eqref{eq:voter.1},
while for $s=2k$ it is embedded in a slightly different way.
More precisely, we set
\begin{equation}
\label{eq:voter.3}
\ffs{2k-1} ~= \left\{
 \begin{array}[h]{ll}
   \frac14\left( \ees{2k-1}+\ees{2k}  +\ees{4k-1}+\ees{4k}\right) &\mbox{if $s\ne2k$}\\[0.5ex]
   \frac14\left( \ees{2k-2}+\ees{2k+1}+\ees{4k-1}+\ees{4k}\right) &\mbox{if $s=2k$.}
 \end{array}
\right.
\end{equation}
Finally, the very last voter $v_{2k}$ is embedded in
\begin{equation}
\label{eq:voter.4}
\ffs{2k} ~=~ \frac14\left( \ees{1}+\ees{2}+\ees{4k-1}+\ees{4k}\right).
\end{equation}
Equations \eqref{eq:voter.1}--\eqref{eq:voter.4} define $\ffs{j}$ for all voters $v_j$ with
$1\le j\le2k$.
This completes the description of the Euclidean representation $\ffz$ of the voters.

We note that the location $\ffs{s}$ of voter $v_s$ has been specified, but will be 
irrelevant for our further arguments.
We will show that $\ffz$ and $\eez$ constitute a correct Euclidean representation 
for the $2k-1$ voters in $\{v_1,\ldots,v_{2k}\}\backslash\{v_s\}$ together with all 
$4k$ alternatives $1,2,\ldots,4k$.
In other words, the deletion of voter $v_s$ from profile $\ppp_k$ yields a one-dimensional 
Euclidean profile, which completes the proof of property~(b) in Theorem~\ref{th:main.1}.
To this end, the following lemma will be established in Section~\ref{sec:correct}.

\begin{lemma}
\label{le:main.(b)}
For all $r$ and $s$ with $1\le r\ne s\le2k$, the Euclidean representation $\eez$ and 
$\ffz$ correctly represents the preferences of voter~$v_r$.
\end{lemma}

The correctness of Lemma~\ref{le:main.(b)} for the small profiles $\ppp_k$ with $k\in\{2,3,4\}$
can easily be verified by a computer program (or by a human prover through tedious case distinctions). 
Hence we will from now on assume that
\begin{equation}
\label{eq:k-greater-4}
k\ge5.
\end{equation}
This assumption will considerably shorten and simplify our arguments.
Note furthermore that the proof of our main result in Theorem~\ref{th:main.2} is not
touched by this assumption, as it builds on the profiles $\ppp_k$ for which $k$ is large 
and tends to infinity.

\section{A collection of technical results}
\label{sec:technical}
In this section we state five technical lemmas.
Lemmas~\ref{le:tech.1} and \ref{le:tech.2} summarize a number of useful identities, 
and will serve as reference tables in our later analysis.
Lemmas~\ref{le:difference} through~\ref{le:tech.ineq} state important 
inequalities that will be central to our proofs.
Throughout we assume that $k\ge5$ according to (\ref{eq:k-greater-4}).

\begin{lemma}
\label{le:tech.1}
For $1\le s\le2k$, the Euclidean embedding $\eez$ satisfies the following.
\begin{subequations}
\begin{alignat}{3}
\label{eq:aux.1a} \ees{2} ~=~ &4k-2s+1  \\[0.5ex]
\label{eq:aux.1b} \ees{3} ~=~ &4k-2s+3  \\[0.5ex]
\label{eq:aux.1c} \ees{4} ~=~ & \left\{
    \begin{array}[h]{ll}
       4k-4s+8 &~\mbox{\rm if $s\in\{1,2\}$}\\[0.5ex]
       8k-4s+8 &~\mbox{\rm if $s\ge3$.}
    \end{array}
  \right.  
\end{alignat}
\end{subequations}
Furthermore for $s\in\{1,2\}$, the embedding $\eez$ satisfies the following.
\begin{subequations}
\begin{alignat}{3}
\label{eq:aux.2a} \ees{2k-2}-\ees{2k-3} ~=~ &4k-2s-7 \\[0.5ex]
\label{eq:aux.2b} \ees{2k-4}-\ees{2k-5} ~=~ &4k-2s-11 
\end{alignat}
\end{subequations}
\end{lemma}
\proof
These statements follow by straightforward calculations from 
\eqref{eq:alter.0}--\eqref{eq:alter.4}.
\qed

\begin{lemma}
\label{le:tech.2}
If (a) $1\le i\le k-1$ and $s\ne2i-1$, or if (b) $i=k$ and $s\notin\{2k-1,2k\}$, the 
following holds:
\begin{subequations}
\begin{alignat}{3}
\label{eq:aux.4a} &\ees{2k+2i}-\ees{2k+2i-1} ~=~ \ees{2i}-\ees{2i-1}+2 
\end{alignat}
\end{subequations}
If (c) $1\le i\le k-1$ and $s\ne 2i$, the following holds:
\addtocounter{equation}{-1}
\begin{subequations}
\addtocounter{equation}{1}
\begin{alignat}{3}
\label{eq:aux.4b} &\ees{2k+2i}-\ees{2k+2i-1} ~=~ \ees{2i+2}-\ees{2i+1}-2
\end{alignat}
\end{subequations}
\end{lemma}
\proof
We distinguish five cases.
The first case assumes $s=2i-1$.
In the setting of the lemma, this case can only occur under (c) with $1\le i\le k-1$.
Then \eqref{eq:alter.4} yields $\ees{2k+2i}-\ees{2k+2i-1}=1$,
while \eqref{eq:alter.2} yields $\ees{2i+2}-\ees{2i+1}=3$.
This implies the desired equality \eqref{eq:aux.4b} for this first case.

The second case assumes $s=2i$.
In the setting of the lemma, this case can only occur under (a) with $1\le i\le k-1$.
Then \eqref{eq:alter.4} yields $\ees{2k+2i}-\ees{2k+2i-1}=4k-1$,
while \eqref{eq:alter.2} yields $\ees{2i}-\ees{2i-1}=4k-3$.
This implies the desired equality \eqref{eq:aux.4a}.

The third case assumes $i=k$.
In the setting of the lemma, this case can only occur under (b) with $1\le s\le 2k-2$.
Then \eqref{eq:alter.6} and \eqref{eq:aux.1a} yield $\ees{4k}-\ees{4k-1}=4k-2s-1$,
while \eqref{eq:alter.2} yields $\ees{2k}-\ees{2k-1}=4k-2s-3$.
This implies the desired equality \eqref{eq:aux.4a}.

In the remaining cases we always have $s\notin\{2i-1,2i\}$.
The fourth case assumes that $1\le i\le k-1$ and that $s=2\ell-1$ is odd,
where $1\le\ell\le k$ and $\ell\ne i$.
In the setting of the lemma, this case can only occur under (a) and (c).
Then \eqref{eq:alter.4} yields
\begin{eqnarray}
\label{eq:uuu.1}
\ees{2k+2i}-\ees{2k+2i-1} &=& 4(i-\ell)+1 \bmod 4k,
\end{eqnarray}
while \eqref{eq:alter.2} yields $\ees{2i}-\ees{2i-1}=4(i-\ell)-1 \bmod 4k$.
Since $i-\ell\ne0$, these two equations together yield \eqref{eq:aux.4a}.
Furthermore, \eqref{eq:alter.2} yields $\ees{2i+2}-\ees{2i+1}= 4(i-\ell)+3 \bmod 4k$,
which together with \eqref{eq:uuu.1} gives \eqref{eq:aux.4b}.

The fifth case assumes that $1\le i\le k-1$ and that $s=2\ell$ is even,
where $1\le\ell\le k$ and $\ell\ne i$.
In the setting of the lemma, this case can only occur under (a) and (c).
Then \eqref{eq:alter.4} yields
\begin{eqnarray}
\label{eq:uuu.2}
\ees{2k+2i}-\ees{2k+2i-1} &=& 4(i-\ell)-1 \bmod 4k,
\end{eqnarray}
while \eqref{eq:alter.2} yields $\ees{2i}-\ees{2i-1}=4(i-\ell)-3 \bmod 4k$.
Since $i-\ell\ne0$, these two statements together imply \eqref{eq:aux.4a}.
Finally, \eqref{eq:alter.2} yields $\ees{2i+2}-\ees{2i+1}= 4(i-\ell)+1 \bmod 4k$.
As $i-\ell\ne0$, this equation together with \eqref{eq:uuu.2} yields \eqref{eq:aux.4b}.
This completes the proof.
\qed

\begin{lemma}
\label{le:difference}
For all alternatives $x$ and $y$ with $1\le y\le x\le4k$, the embedding $\eez$ satisfies
the inequality $\ees{x}-\ees{y}\ge x-y$.
\end{lemma}
\proof 
This follows from Lemma~\ref{le:order} and the integrality of $\eez$.
\qed

\begin{lemma}
\label{le:may-19}
All $i$ and $s$ with $1\le i\le2k-1$ and $1\le s\le2k$ satisfy the following inequality.
\begin{equation}
\label{eq:may-19}
\ees{2i+1}-\ees{2i} ~\ge~2.
\end{equation}
\end{lemma}
\proof
For $1\le i\le k-1$, this follows directly from \eqref{eq:alter.1}.
For $k\le i\le2k-1$, this follows from \eqref{eq:alter.3} and \eqref{eq:alter.5} in
combination with Lemma~\ref{le:difference}.
\qed

\begin{lemma}
\label{le:tech.ineq}
All $i$ and $s$ with $1\le i\le k-1$ and $1\le s\le2k$ satisfy the following inequality.
\begin{alignat}{3}
\label{eq:aux.ineq} \ees{2k+2i-1} ~\ge~ \ees{2k}+\ees{2i}+2.
\end{alignat}
\end{lemma}
\proof
The proof is done by induction on $i=1,\ldots,k-1$.
For the inductive base case $i=1$ we distinguish two subcases on the value of $s$.
The first subcase assumes $s\in\{1,2\}$.
Then \eqref{eq:alter.3} and $k\ge5$, together with \eqref{eq:alter.1}, \eqref{eq:aux.2a}, 
\eqref{eq:aux.2b}, and \eqref{eq:aux.1a} yield
\begin{eqnarray*}
\lefteqn{\ees{2k+2i-1}-\ees{2k} ~\ge~ \ees{2k-1}-\ees{4}+2} \\[0.5ex]
&\ge& (\ees{2k-1}-\ees{2k-2}) +(\ees{2k-2}-\ees{2k-3}) \\
 &&    \quad +(\ees{2k-3}-\ees{2k-4}) +(\ees{2k-4}-\ees{2k-5})+2 \\
&=&   2 +(4k-2s-7) +2 +(4k-2s-11) +2 \\[0.5ex]
&=&   8k-4s-12 ~>~ (4k-2s+1)+2 ~=~ \ees{2}+2.
\end{eqnarray*}
The second subcase assumes $s\ge3$.
Then the first line of \eqref{eq:alter.3} together with $k\ge5$,
\eqref{eq:aux.1c}, \eqref{eq:aux.1b} and \eqref{eq:aux.1a} yields
\begin{eqnarray*}
\lefteqn{\ees{2k+2i-1}-\ees{2k} ~=~ \ees{2k-1}-\ees{3}+2} \\[0.5ex]
&\ge& \ees{4}-\ees{3}+2 ~=~ (8k-4s+8)-(4k-2s+3) +2 \\[0.5ex]
&=&   4k-2s+7 ~>~ \ees{2}+2.
\end{eqnarray*}
Summarizing, in both subcases we have established the desired \eqref{eq:aux.ineq}.
This completes the analysis of the inductive base case $i=1$.
Next, let us state the inductive assumption as
\begin{eqnarray}
\label{eq:vvv.1}
\ees{2k+2i-3} &\ge& \ees{2k}+\ees{2i-2}+2.
\end{eqnarray}
In the inductive step, we will use the following consequence of \eqref{eq:alter.3}:
\begin{eqnarray}
\label{eq:vvv.2}
\ees{2k+2i-1}-\ees{2k+2i-2} &\ge& \ees{2k+2i-3}-\ees{2i+2}+2.
\end{eqnarray}
Furthermore, by \eqref{eq:alter.4} the left hand side of the following inequality equals
$(4i-2s-5\bmod4k)$, while by \eqref{eq:alter.2} its right hand side equals $(4i-2s-3\bmod4k)-2$.
This implies
\begin{eqnarray}
\label{eq:vvv.3}
\ees{2k+2i-2}-\ees{2k+2i-3} &\ge& \ees{2i}-\ees{2i-1}-2.
\end{eqnarray}
Adding up \eqref{eq:vvv.1}, \eqref{eq:vvv.2} and \eqref{eq:vvv.3}, and rearranging and
simplifying the resulting inequality yields
\begin{eqnarray*}
\lefteqn{\ees{2k+2i-1}-\ees{2k}-\ees{2i}-2} \\[0.5ex] 
&\ge& \ees{2k+2i-3}-\ees{2i+2} +\ees{2i-2}-\ees{2i-1} \\[0.5ex] 
&\ge& (2k+2i-3)-(2i+2)-2 ~=~ 2k-7 ~>~ 0.
\end{eqnarray*}
Here we used Lemma~\ref{le:difference} to bound $\ees{2k+2i-3}-\ees{2i+2}$,
and we used \eqref{eq:alter.1} to get rid of $\ees{2i-2}-\ees{2i-1}$.
As this implies \eqref{eq:aux.ineq}, the inductive argument is complete.
\qed

\section{Correctness of the Euclidean representations}
\label{sec:correct}
In this section we prove Lemma~\ref{le:main.(b)}.
Hence, let us fix two arbitrary voters  $v_r$ and $v_s$ with $r\ne s$.
We recall that by Lemma~\ref{le:order} the Euclidean representation $\eez$ embeds
the alternatives $1,\ldots,4k$ in increasing order from left to right.
Our goal is to show that any two alternatives $x$ and $y$ with $x\succ_r y$ that are
consecutive in the preference order of voter $v_r$ satisfy
\begin{subequations}
\begin{alignat}{3}
\label{eq:Euc.1a} &2\ffs{r} ~<~ &\ees{x}+\ees{y} &\mbox{\qquad whenever $x<y$} \\[0.5ex]
\label{eq:Euc.1b} &2\ffs{r} ~>~ &\ees{x}+\ees{y} &\mbox{\qquad whenever $x>y$.}
\end{alignat}
\end{subequations}
By our construction, all preference orders in profile $\ppp_k$ contain long monotone
(increasing or decreasing) runs of alternatives.
By Proposition~\ref{pr:Euclidean.2} it will therefore be sufficient to establish
\eqref{eq:Euc.1a} and \eqref{eq:Euc.1b} at the few turning points where the
preference order of voter $v_r$ changes its monotonicity behavior.
We stress that the first pair of alternatives in every preference order forms a
turning point by default.

The remaining argument is split into four cases that will be handled in the 
following four sections.
Sections~\ref{ssec:1} and~\ref{ssec:2} treat the cases with odd  $r$, while
Sections~\ref{ssec:3} and~\ref{ssec:4} treat the cases with even $r$.

\subsection{The cases with odd r (with a single exception)}
\label{ssec:1}
In this section we consider the cases with odd $r=2i-1$ for $1\le i\le k$, and
with $s\ne2i-1$.
If $i=k$ (and hence $r=2k-1$) then we additionally assume $s\ne2k$; the remaining case 
with $i=k$ and $s=2k$ will be settled in the next section.  
Note that in the cases under current consideration, the value $\ffs{2i-1}$ is given 
by \eqref{eq:voter.1}.
Furthermore \eqref{eq:aux.4a} in Lemma~\ref{le:tech.2} yields 
\begin{alignat}{3}
\label{eq:xxx.0} \ees{2k+2i}+\ees{2i-1} ~=~ \ees{2i} +\ees{2k+2i-1}+2.
\end{alignat}
In order to prove \eqref{eq:Euc.1a} and \eqref{eq:Euc.1b} for the preference orders in 
\eqref{eq:hua.odd} and \eqref{eq:hua.2k-1}, it is sufficient to establish the following 
six inequalities for the turning points.
\begin{subequations}
\begin{alignat}{3}
\label{eq:xxx.1a} &2\ffs{2i-1} ~>~ &&\ees{2k+2i-2}+\ees{2k+2i-3}\\[0.5ex]
\label{eq:xxx.1b} &2\ffs{2i-1} ~<~ &&\ees{2i+1}   +\ees{2k+2i-1}\\[0.5ex]
\label{eq:xxx.1c} &2\ffs{2i-1} ~>~ &&\ees{2k+2i-1}+\ees{2i}     \\[0.5ex]
\label{eq:xxx.1d} &2\ffs{2i-1} ~<~ &&\ees{2i-1}   +\ees{2k+2i}  \\[0.5ex]
\label{eq:xxx.1e} &2\ffs{2i-1} ~>~ &&\ees{2k+2i}  +\ees{2i-2}   \\[0.5ex]
\label{eq:xxx.1f} &2\ffs{2i-1} ~<~ &&\ees{1}      +\ees{2k+2i+1} 
\end{alignat}
\end{subequations}
Note that for $i=1$ the inequality in \eqref{eq:xxx.1e} vanishes as piece $Y_1$ is empty, 
and that for $i=k$ inequality \eqref{eq:xxx.1f} vanishes as piece $Z_k$ is empty.
We use \eqref{eq:voter.1} or the first line of \eqref{eq:voter.3} together with \eqref{eq:xxx.0}, 
and rewrite the common left hand side of all inequalities \eqref{eq:xxx.1a}--\eqref{eq:xxx.1f} as
\begin{alignat}{3}
2\ffs{2i-1} =~ &\frac12\left( \ees{2i-1}+\ees{2i}+\ees{2k+2i-1}+\ees{2k+2i}\right)  \nonumber\\[0.5ex]
            =~ &\ees{2i}+\ees{2k+2i-1}+1 ~=~ \ees{2i-1}+\ees{2k+2i}-1. \label{eq:xxx.2}
\end{alignat}
For \eqref{eq:xxx.1a}, we distinguish two subcases.
The first subcase assumes $i\le k-1$.
We compute by using \eqref{eq:xxx.2}, \eqref{eq:alter.3} with $s\ne2i-1$,
and \eqref{eq:alter.1} that
\begin{eqnarray*}
\lefteqn{2\ffs{2i-1}-\ees{2k+2i-2}-\ees{2k+2i-3}}  \\[0.5ex]
&=& (\ees{2i}+\ees{2k+2i-1}+1) -\ees{2k+2i-2}-\ees{2k+2i-3} \\[0.5ex]
&=& \ees{2i}+1-\ees{2i+1}+2 ~=~ 1 ~>~ 0.
\end{eqnarray*}
The second subcase deals with the remaining case $i=k$.
We use \eqref{eq:xxx.2}, the first line in \eqref{eq:alter.5}, and 
Lemma~\ref{le:order} to compute
\begin{eqnarray*}
\lefteqn{2\ffs{2i-1}-\ees{2k+2i-2}-\ees{2k+2i-3}}  \\[0.5ex]
&=& (\ees{2k}+\ees{4k-1}+1) -\ees{4k-2}-\ees{4k-3} \\[0.5ex]
&=& (\ees{2k}+1)+(-\ees{2}+2) ~=~ (\ees{2k}-\ees{2})+3 ~>~ 0. 
\end{eqnarray*}
For \eqref{eq:xxx.1b}, we compute by using \eqref{eq:xxx.2} and \eqref{eq:may-19} that
\begin{eqnarray*}
\lefteqn{2\ffs{2i-1}-\ees{2i+1} -\ees{2k+2i-1}}  \\[0.5ex]
&=& (\ees{2i}+\ees{2k+2i-1}+1) -\ees{2i+1} -\ees{2k+2i-1} ~<~ 0.
\end{eqnarray*}
For \eqref{eq:xxx.1c}, we compute by using \eqref{eq:xxx.2} that
\begin{eqnarray*}
\lefteqn{2\ffs{2i-1}-\ees{2k+2i-1}-\ees{2i}}  \\[0.5ex]
&=& (\ees{2i}+\ees{2k+2i-1}+1) -\ees{2k+2i-1}-\ees{2i} ~=~ 1 ~>~ 0.
\end{eqnarray*}
For \eqref{eq:xxx.1d}, we compute by using \eqref{eq:xxx.2} that
\begin{eqnarray*}
\lefteqn{2\ffs{2i-1}-\ees{2i-1} -\ees{2k+2i}}  \\[0.5ex]
&=& (\ees{2i-1}+\ees{2k+2i}-1) -\ees{2i-1} -\ees{2k+2i} ~=~ -1 ~<~ 0.
\end{eqnarray*}
For \eqref{eq:xxx.1e} with $i\ge2$, we compute by using \eqref{eq:xxx.2} 
and \eqref{eq:alter.1} that
\begin{eqnarray*}
\lefteqn{2\ffs{2i-1}-\ees{2k+2i}  -\ees{2i-2}}  \\[0.5ex]
&=& (\ees{2i-1}+\ees{2k+2i}-1) -\ees{2k+2i}  -\ees{2i-2} ~=~ 1 ~>~ 0.
\end{eqnarray*}

\bigskip
It remains to prove inequality \eqref{eq:xxx.1f} which takes more effort.
Since \eqref{eq:xxx.1f} vanishes for $i=k$, we may assume $i\le k-1$.
We first use \eqref{eq:xxx.2} and \eqref{eq:alter.0} to derive
\begin{eqnarray}
\lefteqn{2\ffs{2i-1} -\ees{1} -\ees{2k+2i+1}} \nonumber \\[0.5ex]
&=& \ees{2i-1}-1 - (\ees{2k+2i+1}-\ees{2k+2i}). \label{eq:xxx.3}
\end{eqnarray}
Our goal is to show that the value in \eqref{eq:xxx.3} is strictly negative, 
and for this we branch into three subcases.
The first subcase assumes $i\le k-2$.
We use \eqref{eq:alter.3}, \eqref{eq:aux.ineq}, and Lemma~\ref{le:order} to compute
\begin{eqnarray*}
\lefteqn{\ees{2i-1}-1 - (\ees{2k+2i+1}-\ees{2k+2i})}   \\[0.5ex]
   &\le& \ees{2i-1}-1 - (\ees{2k+2i-1}-\ees{2i+4}+2)  \\[0.5ex]
   &\le& \ees{2i-1}-3+\ees{2i+4} - (\ees{2k}+\ees{2i}+2) \\[0.5ex]
   &=&   (\ees{2i+4}-\ees{2k}) +(\ees{2i-1}-\ees{2i}) -5 ~<~ 0.
\end{eqnarray*}
The second subcase assumes $i=k-1$ and $s\ne2k$.
We use \eqref{eq:alter.5}, \eqref{eq:aux.ineq}, and Lemma~\ref{le:order} to compute
\begin{eqnarray*}
\lefteqn{\ees{2i-1}-1 - (\ees{2k+2i+1}-\ees{2k+2i})}   \\[0.5ex]
   &=& \ees{2k-3}-1 - (\ees{4k-1}-\ees{4k-2}) \\[0.5ex]
   &=& \ees{2k-3}-1 - (\ees{4k-3}-\ees{2}+2) \\[0.5ex]
 &\le& \ees{2k-3}-3+\ees{2} - (\ees{2k}+\ees{2k-2}+2) \\[0.5ex]
   &=& (\ees{2k-3}-\ees{2k})+(\ees{2}-\ees{2k-2})-5 ~<~ 0.
\end{eqnarray*}
The third and last subcase assumes $i=k-1$ and $s=2k$.
We use the second line of \eqref{eq:alter.5}, the first line of \eqref{eq:alter.3}, 
inequality \eqref{eq:aux.ineq}, equation \eqref{eq:alter.1}, and Lemma~\ref{le:order} to compute
\begin{eqnarray*}
\lefteqn{\ees{2i-1}-1 - (\ees{2k+2i+1}-\ees{2k+2i})}   \\[0.5ex]
   &=& \ees{2k-3}-1 - (\ees{4k-1}-\ees{4k-2}) \\[0.5ex]
   &=& \ees{2k-3}-1 - (\ees{4k-3}-\ees{2k+1}+2) \\[0.5ex]
   &=& \ees{2k-3}-3+\ees{2k+1} - (\ees{4k-4}+\ees{4k-5}-\ees{2k-1}+2) \\[0.5ex]
 &\le& \ees{2k-3}-5+\ees{2k+1}-\ees{4k-4}\\
    && \quad +\ees{2k-1}-(\ees{2k}+\ees{2k-4}+2) \\[0.5ex]
   &=& (\ees{2k+1}-\ees{4k-4}) + (\ees{2k-1}-\ees{2k}) -5 ~<~ 0.
\end{eqnarray*}
As \eqref{eq:xxx.3} is strictly negative in each of the three subcases, the proof 
of \eqref{eq:xxx.1f} is complete.
The Euclidean representation $\eez$ and $\ffz$ correctly represents the preferences 
of voter $v_r$.

\subsection{The exceptional case with odd r}
\label{ssec:2}
In this section we consider the exceptional case $i=k$ (and hence $r=2k-1$) 
under $s=2k$, which has been left open in the preceding section.
In this exceptional case, the embedding $\ffs{2k-1}$ is given by the second option
in formula \eqref{eq:voter.3}. 
Furthermore, \eqref{eq:alter.6} and \eqref{eq:alter.1} yield
\begin{eqnarray*}
\ees{4k}-\ees{4k-1} &=& \ees{2k+1} -\ees{2k-2}-2. 
\end{eqnarray*}
Altogether this leads to
\begin{alignat}{3}
2\ffs{2k-1} =~ &\frac12\left( \ees{2k-2}+\ees{2k+1}+\ees{4k-1}+\ees{4k}\right)  \nonumber\\[0.5ex]
           =~ &\ees{2k-2}+\ees{4k}+1 ~=~ \ees{2k+1}+\ees{4k-1}-1. \label{eq:xxx.4}
\end{alignat}
As inequality \eqref{eq:xxx.1f} vanishes for $i=k$, our goal in this section is to 
establish the five inequalities \eqref{eq:xxx.1a}--\eqref{eq:xxx.1e} for $i=k$ and $s=2k$.
For \eqref{eq:xxx.1a}, we compute by using \eqref{eq:xxx.4} and \eqref{eq:alter.5} that
\begin{eqnarray*}
\lefteqn{2\ffs{2k-1}-\ees{4k-2}-\ees{4k-3}}  \\[0.5ex]
&=& (\ees{2k+1}+\ees{4k-1}-1) -\ees{4k-2}-\ees{4k-3} \\[0.5ex]
&=& \ees{2k+1}-\ees{4k-3}-1+(\ees{4k-3}-\ees{2k+1}+2) ~=~ 1 ~>~ 0.
\end{eqnarray*}
For \eqref{eq:xxx.1b}, we compute by using \eqref{eq:xxx.4} that
\begin{eqnarray*}
\lefteqn{2\ffs{2k-1}-\ees{2k+1} -\ees{4k-1}}  \\[0.5ex]
&=& (\ees{2k+1}+\ees{4k-1}-1) -\ees{2k+1} -\ees{4k-1} ~=~ -1 ~<~ 0.
\end{eqnarray*}
For \eqref{eq:xxx.1c}, we compute by using \eqref{eq:xxx.4} and \eqref{eq:may-19} that
\begin{eqnarray*}
\lefteqn{2\ffs{2k-1}-\ees{4k-1}-\ees{2k}}  \\[0.5ex]
&=& (\ees{2k+1}+\ees{4k-1}-1) -\ees{4k-1}-\ees{2k} ~>~ 0.
\end{eqnarray*}
For \eqref{eq:xxx.1d}, we compute by using \eqref{eq:xxx.4} and \eqref{eq:alter.1} that
\begin{eqnarray*}
\lefteqn{2\ffs{2k-1}-\ees{2k-1} -\ees{4k}}  \\[0.5ex]
&=& (\ees{2k-2}+\ees{4k}+1) -\ees{2k-1} -\ees{4k} ~=~ -1 ~<~ 0.
\end{eqnarray*}
For \eqref{eq:xxx.1e}, we compute by using \eqref{eq:xxx.4} that 
\begin{eqnarray*}
\lefteqn{2\ffs{2k-1}-\ees{4k}  -\ees{2k-2}}  \\[0.5ex]
&=& (\ees{2k-2}+\ees{4k}+1) -\ees{4k}  -\ees{2k-2} ~=~ 1 ~>~ 0.
\end{eqnarray*}
This completes the analysis of the exceptional case with odd $r$.
Also in this case, the representation $\eez$ and $\ffz$ correctly represents the 
preferences of the considered voter.

\subsection{The cases with even r (with a single exception)}
\label{ssec:3}
In this section we consider the cases with even $r=2i$ for $1\le i\le k-1$, and with $s\ne2i$.
The remaining case $r=2k$ will be settled in the next section.
Note that in the cases under consideration, the value $\ffs{2i}$ is given by \eqref{eq:voter.2}.
Furthermore \eqref{eq:aux.4b} in Lemma~\ref{le:tech.2} yields
\begin{alignat}{3}
\label{eq:yyy.0} &\ees{2i+1}+\ees{2k+2i} ~=~ \ees{2i+2}+\ees{2k+2i-1}-2.
\end{alignat}
In order to prove \eqref{eq:Euc.1a} and \eqref{eq:Euc.1b} for the preference orders 
in \eqref{eq:hua.even}, it is sufficient to establish the following four inequalities 
for the turning points.
\begin{subequations}
\begin{alignat}{3}
\label{eq:yyy.1a} &2\ffs{2i} ~>~ &&\ees{2k+2i-2}+\ees{2k+2i-3} \\[0.5ex]
\label{eq:yyy.1b} &2\ffs{2i} ~<~ &&\ees{2i+2}  +\ees{2k+2i-1} \\[0.5ex]
\label{eq:yyy.1c} &2\ffs{2i} ~>~ &&\ees{2k+2i}  +\ees{2i+1}   \\[0.5ex]
\label{eq:yyy.1d} &2\ffs{2i} ~<~ &&\ees{1}     +\ees{2k+2i+1} 
\end{alignat}
\end{subequations}
We use the definition of $\ffs{2i}$ in \eqref{eq:voter.2} together with \eqref{eq:yyy.0} to 
rewrite the common left hand side of all inequalities \eqref{eq:yyy.1a}--\eqref{eq:yyy.1d} as
\begin{alignat}{3}
2\ffs{2i} =~ &\frac12\left(\ees{2i+1}+\ees{2i+2}+\ees{2k+2i-1}+\ees{2k+2i}\right)  \nonumber\\[0.5ex]
          =~ &\ees{2i+1}+\ees{2k+2i}+1 ~=~ \ees{2i+2}+\ees{2k+2i-1}-1. \label{eq:yyy.2}
\end{alignat}
For \eqref{eq:yyy.1a}, we compute by using \eqref{eq:yyy.2} and \eqref{eq:alter.3} that 
\begin{eqnarray*}
\lefteqn{2\ffs{2i}-\ees{2k+2i-2}-\ees{2k+2i-3}}  \\[0.5ex]
  &=& (\ees{2i+2}+\ees{2k+2i-1}-1) -\ees{2k+2i-2}-\ees{2k+2i-3} \\[0.5ex]
&\ge& \ees{2i+2}-\ees{2k+2i-3}-1+(\ees{2k+2i-3}-\ees{2i+2}+2) =1 >0.
\end{eqnarray*}
For \eqref{eq:yyy.1b}, we compute by using \eqref{eq:yyy.2} that 
\begin{eqnarray*}
\lefteqn{2\ffs{2i}-\ees{2i+2}-\ees{2k+2i-1}}  \\[0.5ex]
  &=& (\ees{2i+2}+\ees{2k+2i-1}-1) -\ees{2i+2}-\ees{2k+2i-1} = -1 < 0.
\end{eqnarray*}
For \eqref{eq:yyy.1c}, we compute by using \eqref{eq:yyy.2} that
\begin{eqnarray*}
\lefteqn{2\ffs{2i}-\ees{2k+2i}-\ees{2i+1}}  \\[0.5ex]
  &=& (\ees{2i+1}+\ees{2k+2i}+1) -\ees{2k+2i}-\ees{2i+1} = 1 > 0.
\end{eqnarray*}

\bigskip
It remains to prove inequality \eqref{eq:yyy.1d} which takes a considerable amount of work.
We branch into three subcases.
The first subcase assumes $1\le i\le k-2$.
Then Lemma~\ref{le:order} implies $\ees{2i+4}\le\ees{2k}$.
We use \eqref{eq:yyy.2}, \eqref{eq:alter.0}, \eqref{eq:alter.3}, \eqref{eq:aux.ineq} and 
\eqref{eq:alter.1} to derive
\begin{eqnarray*}
\lefteqn{2\ffs{2i} -\ees{1} -\ees{2k+2i+1}} \\[0.5ex]
  &=& (\ees{2i+1}+\ees{2k+2i}+1)  -\ees{2k+2i+1} \\[0.5ex]
&\le& \ees{2i+1}+1 - (\ees{2k+2i-1}-\ees{2i+4}+2) \\[0.5ex]
&\le& \ees{2i+1}+\ees{2i+4}-1 -(\ees{2k}+\ees{2i}+2) \\[0.5ex]
  &=& \ees{2i+4}-\ees{2k}-1 ~<~ 0.
\end{eqnarray*}

The second subcase assumes $i=k-1$ and $s\ne2k$.
For proving \eqref{eq:yyy.1d}, we compute by using \eqref{eq:yyy.2}, \eqref{eq:alter.0},
\eqref{eq:alter.5}, \eqref{eq:aux.ineq} and \eqref{eq:alter.1} that
\begin{eqnarray*}
\lefteqn{2\ffs{2k-2} -\ees{1} -\ees{4k-1}}   \\[0.5ex]
  &=& (\ees{2k-1}+\ees{4k-2}+1)  -\ees{4k-1} \\[0.5ex]
  &=& \ees{2k-1}+1 -(\ees{4k-3}-\ees{2}+2)   \\[0.5ex]
&\le& \ees{2k-1}+\ees{2}-1 -(\ees{2k}+\ees{2k-2}+2) \\[0.5ex]
  &=& \ees{2}-\ees{2k}-1 ~<~ 0.
\end{eqnarray*}

The third and last subcase finally assumes $i=k-1$ and $s=2k$.
We start the analysis by deriving a number of auxiliary equations and inequalities.
First we determine $\ees{3}=3$ from \eqref{eq:aux.1b}, and compute by using 
\eqref{eq:alter.3} and \eqref{eq:alter.1} that
\begin{eqnarray}
\nonumber \lefteqn{\ees{2k+1} -\ees{2k-2}}  \\[0.5ex]
\nonumber         &=& (\ees{2k+1}-\ees{2k}) +(\ees{2k}-\ees{2k-1}) +(\ees{2k-1}-\ees{2k-2})\\[0.5ex]
\label{eq:yyy.3a} &=& (\ees{2k-1}-\ees{3}+2) +(\ees{2k}-\ees{2k-1}) +2 ~~=~~ \ees{2k}+1.
\end{eqnarray}
Next, we use \eqref{eq:aux.4b} to derive 
\begin{eqnarray}
\label{eq:yyy.3b} 
\ees{2k-2}-\ees{2k-3}-2 &=& \ees{4k-4}-\ees{4k-5}.
\end{eqnarray}
We express $\ees{4k-3}$ once by the first line of \eqref{eq:alter.3} and once by 
the second line of \eqref{eq:alter.5}, which by equating yields 
\begin{eqnarray}
\lefteqn{\ees{4k-4}+\ees{4k-5}-\ees{2k-1}+2} \nonumber \\[0.5ex] 
&=& \ees{2k+1}-2 +\ees{4k-1}-\ees{4k-2}.
\label{eq:yyy.3c}
\end{eqnarray}
Next, we add up \eqref{eq:yyy.3a}, \eqref{eq:yyy.3b}, \eqref{eq:yyy.3c} and
rearrange the result to derive
\begin{eqnarray}
\lefteqn{\ees{2k-1}+\ees{4k-2}-\ees{4k-1}+1} \nonumber \\[0.5ex] 
&=& 2\ees{2k-1} +\ees{2k}+\ees{2k-3}-2\ees{4k-5}.
\label{eq:yyy.3d}
\end{eqnarray}
We compute $\ees{2k}-\ees{2k-1}=4k-3$ and $\ees{2k-2}-\ees{2k-3}=4k-7$ from 
\eqref{eq:alter.2}, and use these together with \eqref{eq:alter.1} to get
\begin{eqnarray}
\nonumber \lefteqn{\ees{2k-1}-\ees{2k-4}+\ees{2k-1}-\ees{2k}} \\[0.5ex] 
\nonumber        &=& (\ees{2k-2}+2)-(\ees{2k-3}-2)-(\ees{2k}-\ees{2k-1}) \\[0.5ex]
\label{eq:yyy.4} &=& 2+(4k-7)+2-(4k-3) ~=~ 0. 
\end{eqnarray}
Now for finally proving \eqref{eq:yyy.1d} in this third and last subcase, we compute by 
using \eqref{eq:yyy.2}, \eqref{eq:alter.0}, \eqref{eq:yyy.3d}, \eqref{eq:aux.ineq}, 
\eqref{eq:yyy.4} and \eqref{eq:alter.1} that
\begin{eqnarray*}
\lefteqn{2\ffs{2k-2} -\ees{1} -\ees{4k-1}}          \\[0.5ex]
  &=& (\ees{2k-1}+\ees{4k-2}+1)  -\ees{4k-1}        \\[0.5ex]
  &=& 2\ees{2k-1} +\ees{2k}+\ees{2k-3}-2\ees{4k-5}  \\[0.5ex]
&\le& 2\ees{2k-1} +\ees{2k}+\ees{2k-3}-2(\ees{2k}+\ees{2k-4}+2) \\[0.5ex]
  &=& \ees{2k-3}-\ees{2k-4}-4 ~=~ -2 ~<~ 0.
\end{eqnarray*}
This completes the proof of inequality \eqref{eq:yyy.1d}.
Summarizing, the representation $\eez$ and $\ffz$ correctly represents the preferences 
of the considered voter $v_r$.

\subsection{The exceptional case with even r}
\label{ssec:4}
In this section we consider the last remaining case with even $r$, where $r=2k$ and $s\ne2k$ holds.
In order to prove \eqref{eq:Euc.1a} and \eqref{eq:Euc.1b} for the preference orders
in \eqref{eq:hua.2k}, it is sufficient to establish the following three inequalities
for the turning points.
\begin{subequations}
\begin{alignat}{3}
\label{eq:zzz.1a} &2\ffs{2k} ~>~ &&\ees{4k-2}+\ees{4k-3} \\[0.5ex]
\label{eq:zzz.1b} &2\ffs{2k} ~<~ &&\ees{2}   +\ees{4k-1} \\[0.5ex]
\label{eq:zzz.1c} &2\ffs{2k} ~>~ &&\ees{4k}  +\ees{1}   
\end{alignat}
\end{subequations}
The definition of $\ffs{2k}$ in \eqref{eq:voter.4} and  \eqref{eq:alter.6} with $s\ne2k$
yield for the common left hand side of \eqref{eq:zzz.1a}--\eqref{eq:zzz.1c} that
\begin{alignat}{3}
2\ffs{2k} =~ &\frac12\left(\ees{1}+\ees{2}+\ees{4k-1}+\ees{4k}\right)  \nonumber\\[0.5ex]
         =~ &\ees{4k}+\ees{1}+1 ~=~ \ees{4k-1}+\ees{2}-1.  \label{eq:zzz.2}
\end{alignat}
For \eqref{eq:zzz.1a}, we compute by using \eqref{eq:zzz.2} and 
\eqref{eq:alter.5} with $s\ne2k$ that
\begin{eqnarray*}
\lefteqn{2\ffs{2k}-\ees{4k-2}-\ees{4k-3}}  \\[0.5ex]
&=& (\ees{4k-1}+\ees{2}-1) -\ees{4k-2}-\ees{4k-3} \\[0.5ex]
&=& \ees{2}-\ees{4k-3}-1 +(\ees{4k-3}-\ees{2}+2) ~=~ 1 ~>~ 0.
\end{eqnarray*}
For \eqref{eq:zzz.1b}, we compute by using \eqref{eq:zzz.2} that
\begin{eqnarray*}
\lefteqn{2\ffs{2k}-\ees{2} -\ees{4k-1}}  \\[0.5ex]
&=& (\ees{4k-1}+\ees{2}-1) -\ees{2} -\ees{4k-1} ~=~ -1 ~<~ 0.
\end{eqnarray*}
For \eqref{eq:zzz.1c}, we compute by using \eqref{eq:zzz.2} that 
\begin{eqnarray*}
\lefteqn{2\ffs{2k}-\ees{4k}  +\ees{1}}  \\[0.5ex]
&=& (\ees{4k}+\ees{1}+1) -\ees{4k}  -\ees{1} ~=~ 1 ~>~ 0.
\end{eqnarray*}
This settles the last case. 
The proof of Lemma~\ref{le:main.(b)} and with it the proof of Theorem~\ref{th:main.1} 
are finally complete.

\section{Conclusions}
\label{sec:conclusion}
We have shown that one-dimensional Euclidean preference profiles can not be characterized 
in terms of finitely many obstructions.
This is similar to the situation of interval graphs, which also can not be characterized
by finitely many obstructions.
For interval graphs, however, we have a full understanding of all the obstructions that are 
\emph{minimal} with respect to vertex deletion; see Lekkerkerker \& Boland \cite{LeBo1962}.
In a similar vein, it would be interesting to determine all the (infinitely many) obstructions 
for one-dimensional Euclidean preferences that are minimal with respect to deletion of 
voters or alternatives.
At the current moment, we have no idea of what these minimal obstructions would look like.

With respect to general $d$-dimensional Euclidean preference profiles, we feel that the 
situation should be analogous to the one-dimensional situation: 
we conjecture that for any fixed value of $d\ge2$, there will be no characterization 
of $d$-dimensional Euclidean profiles through finitely many obstructions.
However, we see no realistic way of generalizing our current approach to the higher-dimensional 
situations, and we leave this as an open problem.
(We remind the reader that in a $d$-dimensional Euclidean preference profile the voters and
alternatives are embedded in $d$-dimensional Euclidean space, so that small distance 
corresponds to strong preference; see for instance Bogomolnaia \& Laslier \cite{BoLa2007}.)

\section*{Acknowledgements}
This research has been supported by COST Action IC1205 on Computational Social Choice.
Jiehua Chen acknowledges support
  by the Studienstiftung des Deutschen Volkes.
Kirk Pruhs is supported 
  in part by NSF grants CCF-1115575, CNS-1253218, CCF-1421508, and an IBM Faculty Award.
Gerhard Woeginger acknowledges support
  by the Zwaartekracht NETWORKS grant of NWO, 
  and by the Alexander von Humboldt Foundation, Bonn, Germany.




\medskip

\begin{thebibliography}{28}

\bibitem{BaHa2011}
{\sc M.A. Ballester and G. Haeringer} (2011).
A characterization of the single-peaked domain.
\emph{Social Choice and Welfare 36}, 305--322.

\bibitem{BaJa2004}
{\sc S. Barber\`a and M.O. Jackson} (2004).
Choosing how to choose:  self-stable majority rules and constitutions.
\emph{Quarterly Journal of Economics 119}, 1011--1048.

\bibitem{BaTr1986}
{\sc J. Bartholdi III and M.A. Trick} (1986).
Stable matching with preferences derived from a psychological model.
\emph{Operations Research Letters 5}, 165--169.

\bibitem{Black1948}
{\sc D. Black} (1948).
On the rationale of group decision-making.
\emph{Journal of Political Economics 56}, 23--34.

\bibitem{BoLa2007}
{\sc A. Bogomolnaia and J.F. Laslier} (2007).
Euclidean preferences.
\emph{Journal of Mathematical Economics 43}, 87--98.

\bibitem{BrJoKi2002}
{\sc S.J. Brams, M.A. Jones, and D.M. Kilgour} (2002).
Single-peakedness and disconnected coalitions.
\emph{Journal of Theoretical Politics 14}, 359--383.  

\bibitem{BrChWo2013a}
{\sc R. Bredereck, J. Chen, and G.J. Woeginger} (2013).
A characterization of the single-crossing domain. 
\emph{Social Choice and Welfare 41}, 989--998.

\bibitem{BrChWo2013b}
{\sc R. Bredereck, J. Chen, and G.J. Woeginger} (2013).
Are there any nicely structured preference profiles nearby? 
In \emph{Proceedings of the 23rd International Joint Conference on Artificial Intelligence (IJCAI'2013)}, 62--68.

\bibitem{Coombs1964}
{\sc C. Coombs} (1964).
\emph{A Theory of Data}.
John Wiley and Sons, New York.

\bibitem{Demange1994}
{\sc G. Demange} (1994).
Intermediate preferences and stable coalition structures.
\emph{Journal of Mathematical Economics 23}, 45--58.

\bibitem{DiSt1974}
{\sc P.A. Diamond and J.E. Stiglitz} (1974). 
Increases in risk and in risk aversion. 
\emph{Journal of Economic Theory 8}, 337--360.

\bibitem{DoFa1994}
{\sc J. Doignon and J. Falmagne} (1994).
A polynomial time algorithm for unidimensional unfolding representations.
\emph{Journal of Algorithms 16}, 218--233.

\bibitem{ElFaSl2012}
{\sc E. Elkind, P. Faliszewski, and A.M. Slinko} (2012).
Clone structures in voters' preferences.
In \emph{Proceedings of the 13th ACM Conference on Electronic Commerce (EC'12)}, 496--513.  

\bibitem{ElLa2014}
{\sc E. Elkind and M. Lackner} (2014).
On detecting nearly structured preference profiles. 
In \emph{Proceedings of the 28th AAAI Conference on Artificial Intelligence (AAAI'2014)}, 661--667 .

\bibitem{EpPl1998}
{\sc D. Epple and G.J. Platt} (1998).
Equilibrium and local redistribution in an urban economy when households differ in both preferences and incomes.
\emph{Journal of Urban Economics 43}, 23--51.

\bibitem{EsLaOz2008}
{\sc B. Escoffier, J. Lang, and M. {\"O}zt{\"u}rk} (2008).
Single-peaked consistency and its complexity.
In \emph{Proceedings of the 18th European Conference on Artificial Intelligence (ECAI'08)}, 366--370. 

\bibitem{FoHa1977}
{\sc S. F\"oldes and P.L. Hammer} (1977). 
Split graphs. 
In \emph{Proceedings of the Eighth Southeastern Conference on Combinatorics, Graph Theory and Computing}, 311--315.

\bibitem{GaSm1996}
{\sc J.S. Gans and M. Smart} (1996).
Majority voting with single-crossing preferences.
\emph{Journal of Public Economics 59}, 219--237.

\bibitem{Grandmont1978}
{\sc J.-M. Grandmont} (1978).
Intermediate preferences and majority rule.
\emph{Econometrica 46}, 317--330.

\bibitem{HoKoSa1985}
{\sc A.J. Hoffman, A.W.J. Kolen, and M. Sakarovitch} (1985).
Totally-balanced and greedy matrices.
\emph{SIAM Journal on Algebraic Discrete Methods 6}, 721--730.

\bibitem{Hotelling1929}
{\sc H. Hotelling} (1929). 
Stability in competition. 
\emph{Economic Journal 39(153)}, 41--57.

\bibitem{Inada1969}
{\sc K. Inada} (1969).
The simple majority rule.
\emph{Econometrica 37}, 490--506.

\bibitem{Karlin1968}
{\sc S. Karlin} (1968). 
\emph{Total Positivity}. 
Stanford University Press.

\bibitem{Knoblauch2010}
{\sc V. Knoblauch} (2010).
Recognizing one-dimensional {E}uclidean preference profiles.
\emph{Journal of Mathematical Economics 46}, 1--5.  

\bibitem{Kung2006}
{\sc F.-C. Kung} (2006).
An algorithm for stable and equitable coalition structures with public goods.
\emph{Journal of Public Economic Theory 8}, 345--355.

\bibitem{Kuratowski1930}
{\sc K. Kuratowski} (1930).
Sur le probl\`eme des courbes gauches en topologie.
\emph{Fundamenta Mathematicae 15}, 271--283.

\bibitem{LeBo1962}
{\sc C. Lekkerkerker and D. Boland} (1962).
Representation of finite graphs by a set of intervals on the real line.
\emph{Fundamenta Mathematicae 51}, 45--64.

\bibitem{MeRi1981}
{\sc A.H. Meltzer and S.F. Richard} (1981). 
A rational theory of the size of government.
\emph{Journal of Political Economy 89}, 914--927.

\bibitem{Mirrlees1971}
{\sc J.A. Mirrlees} (1971).
An exploration in the theory of optimal income taxation.
\emph{Review of Economic Studies 38}, 175--208.

\bibitem{Moulin1980}
{\sc H. Moulin} (1980).
On strategy-proofness and single peakedness.
\emph{Public Choice 35}, 437--455.

\bibitem{Roberts1977}
{\sc K.W.S. Roberts} (1977).
Voting over income tax schedules.
\emph{Journal of Public Economics 8}, 329--340.

\bibitem{Westhoff1977}
{\sc F.  Westhoff} (1977).
Existence of equilibria in economies with a local public good.
\emph{Journal of Economic Theory 14}, 84--112.

\end{thebibliography}
\end{document}